\documentstyle[12pt]{article}

\def\be{\begin{equation}}
\def\ee{\end{equation}}
\def\ben{$$}
\def\een{$$}
\def\ba{\begin{array}{c}}
\def\ea{\end{array}}
\def\p{\partial}
\begin{document}

\titlepage
\vspace*{1cm}

\begin{center}{\Large \bf
 New perturbation method

 with the matching of wave functions
 }\end{center}

\vspace{10mm}

\begin{center}
Miloslav Znojil \vspace{3mm}

\'{U}stav jadern\'e fyziky AV \v{C}R, 250 68 \v{R}e\v{z}, Czech
Republic\\

e-mail: znojil@ujf.cas.cz, \today

\end{center}

\vspace{5mm}

\section*{Abstract}

We propose a new approach to the Rayleigh-Schr\"{o}dinger
perturbation expansions of bound states in quantum mechanics. We
are inspired by the enormous flexibility of solvable interactions
with several ($N$) discontinuities. Their standard matching
solution is modified and transferred in perturbation regime. We
employ the {\em global} renormalization freedom of the {\em local}
wave functions and derive a compact $N-$dimensional matrix formula
for corrections. In applications, our recipe is shown
non-numerical for all polynomial perturbations of any piece-wise
constant zero order potential.

\vspace{5mm}

\subsubsection*{[KEYWORDS]}

\noindent [Matching method] for \noindent
[Rayleigh-Schr\"{o}dinger corrections] to \noindent [bound states]
within \noindent [perturbation theory (using) symbolic
manipulations] applied to \noindent [piece-wise constant
potentials (with) polynomial perturbations].

\newpage

\section{Introduction}

Rayleigh-Schr\"{o}dinger perturbation theory \cite{Kato} leads to
several popular and efficient numerical approximation methods
\cite{Zinn}. Its construction of observables may also prove
inspiring in the more abstract analysis of their coupling
dependence \cite{Harrell}. The latter role of the perturbative
power series ansatzs has already been emphasized in the classical
monograph by Morse and Feshbach \cite{Morse}. Their presentation
of the Rayleigh-Schr\"{o}dinger formalism contemplates {\em any}
potential $V(r)=V^{(0)}(r) + \lambda\,V^{(1)}(r)$ as defined on a
{\em trivial} square-well background $V^{(0)}(r)$.

In the late sixties the ``mainstream" attention has been shifted
to analytic $V^{(0)}(r)$. People have noticed that a suitable
normalization leads to five-term recurrences and to an enormous
simplification of the construction of the anharmonic oscillators
with $V^{(1)}(r) \sim r^{4}$ \cite{Kunihiro} etc. The history has
been reviewed, e.g., by Simon \cite{Simon}.

An unpleasant obstacle to a broader applicability of perturbative
solutions formed by the power series in $\lambda$ is definitely
the narrow variability of the available analytic zero order
approximations. In three dimensions Newton \cite{Newton} only
lists square well $V^{(0)}(r) \sim (r/L)^{p}, \ p \to \infty$,
harmonic well $V^{(0)}(r) \sim (r/L)^{2}$, Coulomb field
$V^{(0)}(r) \sim (r/L)^{-1}$ and a rather exotic strongly singular
$V^{(0)}(r) \sim (r/L)^{-4}$ \cite{Papp}.

The non-analytic square well seems tedious in comparison but in
the present paper we still return to discontinuous $V^{(0)}(r)$. A
new version of the Rayleigh-Schr\"{o}dinger method will be
proposed. We shall argue that an appropriate ``optimal"
normalization is equally well able to simplify many models
containing $N$ discontinuities in a way which enhances
significantly the flexibility of the above-mentioned
Morse-Feshbach single-well example.

Our conjecture is based on several technical ingredients. Firstly,
we imagine that the contemporary computers shift the borderlines
of feasibility of the initial zero order constructions. In this
sense the standard solvable square well may easily be complemented
not only by its textbook modification of finite depth
\cite{Fluegge} but by virtually any piece-wise constant potential
$V^{(0)}(r)$. This is discussed in Section 2. For definitness we
pick up there the $s-$wave ($\ell=0$) rectangular or step-shaped
example
 \be V_{(N)}^{(0)}(r) = \left\{
\begin{array}{lll}
\infty,&\ \ \ \ r \in (-\infty,L_0) \bigcup (L_{N+1},\infty)\  &\\
 H_j,&\ \ \ \ r \in (L_{j},L_{j+1}),&\ \ \
 \ \ j = 0, 1,  \ldots, N
 \ea \right.
 \label{zeroor}
  \ee
with impenetrable outer barriers $ H_{N+1}=\infty$ and $
H_{-1}=\infty$ and with $N$ discontinuities and $L_0 \geq 0$.
These specifications are just convenient and could easily be
altered.

The study of perturbations of the more and more complicated
solvable potentials (\ref{zeroor}) reveals that the standard use
of an unperturbed basis may become prohibitively cumbersome. The
evaluation of all the necessary Rayleigh-Schr\"{o}dinger overlap
integrals does not seem rewarded by the resulting series
 \be
\ba \psi (x) = \psi^{(0)}(x) + \lambda \ \psi^{(1)}(x) + \lambda^2
\psi^{(2)}(x) + \ldots,
\\
 E = E^{(0)} + \lambda E^{(1)} + \lambda^2 E^{(2)} + \ldots\ .
\ea \label{7}
 \ee
A simpler recipe is asked for. In section 3 we propose, therefore,
a new approach to corrections which circumvents the use of
integrals. We shall see how it combines the continuity of our
Schr\"{o}dinger differential equation on certain finite intervals
with an ease of their mutual matching.

The formal appeal of our new technique lies in its unexpectedly
coherent combination of the matching of perturbed wave functions
$\psi^{(k)}(r)$ with a ``hidden" freedom of their normalization.
The idea transcends its present application and makes the
formalism quite universal. Our matching of perturbation
corrections may be understood as a useful alternative to the
standard textbook recipe even in applications to smooth
potentials. This is discussed in the last Section 4.

\section{Matching method in zero order \label{2}}

We usually expect that a split of a given potential $V(x)$ into a
dominant part $V^{(0)}(x)$ and its perturbation
$\lambda\,V^{(1)}(x)$ simplifies our Schr\"{o}dinger equation in
its unperturbed limit $\lambda \to 0$. Potentials $V^{(0)}(x)$ are
predominantly chosen as harmonic oscillators. The more complicated
shapes of $V(x)$ can hardly be treated by perturbation expansions
without resort to their discontinuous approximants.

\subsection{The piece-wise constant unperturbed potentials
\label{2.1}}

The first non-trivial $s-$wave step-like example (\ref{zeroor})
with $N=1$, gauge $H_0=0$, energy $E=\beta^2$, step $
H_1=\beta^2-\gamma^2$ and abbreviations $L_0=0$, $L_1=P$, $L_2=Q$
possesses the trivial wave functions
 \ben \psi^{(0)}(x) = \left\{
\begin{array}{ll}
{\beta^{-1}\sin \beta x } ,&\ \ \ \ x \in (0,P)\\ { {\cal N}
\,\gamma^{-1}\sin \gamma (x-Q) } ,&\ \ \ \ x \in (P,Q)\ . \ea
\right.
 \een
Their matching at $P$ fixes $ {\cal N} $ and defines the energies
as roots of the elementary trigonometric equation
 \ben
\gamma\,{{\rm tg} \beta P } = \beta\, {{\rm tg} \gamma ( P-Q) } \
.
 \een
In the less common $N=2$ example with $L_3=R$ let us abbreviate
$E=\beta^2$, $\alpha^2=H_1-\beta^2$ and $\gamma^2=\beta^2-H_2$ and
admit the complex $\alpha$, $\beta$ and $\gamma$ in
 \ben
\psi^{(0)}(x) = \left\{
\begin{array}{ll}
{\beta^{-1}\sin \beta x },&\ \ \ \ x \in (0,P)\\
\alpha^{-1}\left(c_1e^{\alpha(x-P)}-c_2e^{-\alpha(x-P)}\right) ,&\
\ \ \ x \in (P,Q)\\ {{\cal N}\,\gamma^{-1}\sin \gamma (x-Q) },&\ \
\ \ x\in (Q,R)\ . \ea \right.
 \een
In terms of $B = \beta P$, $C = \gamma(Q-R)$, $  A ={\rm
arctg}(\alpha/\beta) $ and $  D ={\rm arctg}(\alpha/\gamma)$ the
spectrum follows from the similar matching condition
\be
e^{\alpha(Q-P)} \cos (B-A) \cos (C+D) - e^{-\alpha(Q-P)} \cos
(B+A) \cos (C-D)=0 . \label{DWAsec}
 \ee
The graphical localization of its zeros is sampled in Figure 1
where we have chosen $P=1,\ Q=2$ and $R=\pi$. With the double-well
choice of $H_2=0$ the Figure displays our double-well secular
determinant at the four different heights of its central barrier.
In principle, these curves range from zero up to the maximal
$k=\beta=\sqrt{H_1}$ where they turn purely imaginary.  Our
picture shows just a small vicinity of the quasi-degenerate
doublet of the two lowest energies. Due to the asymmetry of
$V^{(0)}(x)$ their split only very weakly depends on the repulsive
central core.

\subsection{Trigonometric symbolic manipulations\label{2.2}}

After we move to the higher integers $N$ the assistance of a
computer becomes welcome.
For example, the choice of $N=4$ may mimic a double tunneling.
With $H_0=H_2=H_4=0$ and $H_1=H_3=H>0$, at the energy $E =
\kappa^2=H\,\cos \alpha$ and with abbreviations $L_4=S$, $L_5=T$,
$\delta=\sqrt{H-E} \equiv \sqrt{H} \sin \alpha>0$ and
 \ben F(\alpha)=\sin [(R-Q)\,\kappa(\alpha)] \, \cos
[(T-S)\,\kappa(\alpha)+\alpha]\, \cos [P\,\kappa(\alpha)-\alpha],
\een \ben G(\alpha)= \sin [2\alpha-(R-Q)\,\kappa(\alpha)] \, \cos
[(T-S)\,\kappa(\alpha)-\alpha]\, \cos [P\,\kappa(\alpha)-\alpha]
 \een
we derive the secular equation
 \ben
 {e^{(Q-P)\delta}}
\left[ F(\alpha){e^{(S-R)\delta}} +G(\alpha){e^{-(S-R)\delta}}
\right] = {e^{-(Q-P)\delta}} \left[ F(-\alpha){e^{(S-R)\delta}}
+G(-\alpha){e^{-(S-R)\delta}} \right].
 \een
It is fairly transparent. In our last example with $N=6$ the
triplet of barriers in
 \ben V_{(QW)}(x) = \left\{
\begin{array}{ll} 0,&\ \ \ \ x \in (0,P)\bigcup(Q,R)\bigcup (S,T)\bigcup(U,W)
\\ H,&\ \ \ \ x \in (P,Q)\bigcup(R,S)\bigcup(T,U)\\ \infty,&\ \ \ \ x \in
(-\infty,0)\bigcup (W, \infty) \ea \right.
 \een
requires similar strategy. With the four auxiliary functions
 \ben \ba
 F_1= \cos (P\,\kappa+\alpha)\,
\sin [(R-Q)\,\kappa+2\,\alpha] \, \sin [(T-S)\,\kappa+2\,\alpha]
\, \cos [(W-U)\,\kappa+\alpha],\\
 F_2= \cos (P\,\kappa+\alpha)\,
\sin [(R-Q)\,\kappa+2\,\alpha] \, \sin [(T-S)\,\kappa] \, \cos
[(W-U)\,\kappa-\alpha],\\
 F_3= \cos (P\,\kappa+\alpha)\,
\sin [(R-Q)\,\kappa] \, \sin [(T-S)\,\kappa] \, \cos
[(W-U)\,\kappa+\alpha],\\
 F_4= \cos (P\,\kappa+\alpha)\,
\sin [(R-Q)\,\kappa] \, \sin [(T-S)\,\kappa-2\,\alpha] \, \cos
[(W-U)\,\kappa-\alpha]\ \ea
 \een
of $\alpha$ and $\kappa=\kappa(\alpha) =\kappa(-\alpha) =\sqrt{H}
\cos\alpha\equiv \sqrt{E}>0$ and with the same $\delta
=|\delta|=\sqrt{H-E}\equiv \sqrt{H} \sin\alpha>0$ as above, the
exact secular equation is
 \ben \ba \left[-F_1(\alpha)\,
{e^{2(P+R+T)\delta(\alpha)}} +F_2(\alpha)\,
{e^{2(P+R+U)\delta(\alpha)}} +F_3(\alpha)\,
{e^{2(P+S+T)\delta(\alpha)}} \right. \\ -F_4(\alpha)\,
{e^{2(P+S+U)\delta(\alpha)}} +F_1(-\alpha)\,
{e^{2(Q+S+U)\delta(\alpha)}} -F_2(-\alpha)\,
{e^{2(Q+S+T)\delta(\alpha)}}\\ \left. -F_3(-\alpha)\,
{e^{2(Q+R+U)\delta(\alpha)}} +F_4(-\alpha)\,
{e^{2(Q+R+T)\delta(\alpha)}} \right]\,
e^{[(-P-Q-R-S-T-U)\delta(\alpha)]}=0\ . \ea
 \een
The study of the other similar systems may be guided by this
experience. The longer ansatzs remain tractable by computerized
trigonometric manipulations.

\subsection{The general matching recipe \label{2.3}}

Any Schr\"{o}dinger bound state problem with a piece-wise constant
potential is exactly solvable on each sub-interval $(L_j,
L_{j+1})$. Its two independent solutions are just the
(trigonometric or hyperbolic) sines ${\cal S}_j^{(+)}(x)$ and
cosines ${\cal C}_j^{(+)}(x)$ determined in a unique way by the
left initial conditions,
 \be
 \ba
{\cal C}^{(+)}_{j}(x) \left. \ba
\\
\ea\!\!\!\!\!\!\! \right|_{x=L_j} = 1, \ \ \ \ \ \ \ \p_x {\cal
C}^{(+)}_{j}(x) \left. \ba
\\
\ea\!\!\!\!\!\!\! \right|_{x=L_j} = 0,\\
 {\cal S}^{(+)}_{j}(x) \left. \ba
\\
\ea\!\!\!\!\!\!\! \right|_{x=L_j} = 0, \ \ \ \ \ \ \ \p_x {\cal
S}^{(+)}_{j}(x) \left. \ba
\\
\ea\!\!\!\!\!\!\! \right|_{x=L_j} = 1 .
 \ea \label{inibula}
  \ee
They define the general solution simply as a superposition
 \be \psi^{(0)}(x) = c^{(+)}(j)\,{\cal C}^{(+)}_j(x)
  + d^{(+)}(j)\, {\cal S}^{(+)}_j(x)\ , \ \ \ \ x
\in (L_{j}, L_{j+1}), \ \ \ j =0, 1,  \ldots,N.
 \label{generalula}
 \ee
In the light of the obvious symmetry of our global problem on
$(L_0, L_{N+1})$, one can equally well employ the alternative
ansatz
 \be \psi^{(0)}(x) = c^{(-)}({j+1})\,{\cal C}^{(-)}_{j+1}(x)
  + d^{(-)}({j+1})\, {\cal S}^{(-)}_{j+1}(x)\ , \ \ \ \ x
\in (L_{j}, L_{j+1}), \ \ \
 \label{generaluben}
 \ee
with the $^{(-)}-$superscripted basis defined by the right initial
conditions
 \be
 \ba
{\cal C}^{(-)}_{{j+1}}(x) \left. \ba
\\
\ea\!\!\!\!\!\!\! \right|_{x=L_{j+1}} = 1, \ \ \ \ \ \ \ \p_x
{\cal C}^{(-)}_{{j+1}}(x) \left. \ba
\\
\ea\!\!\!\!\!\!\! \right|_{x=L_{j+1}} = 0,\\
 {\cal S}^{(-)}_{{j+1}}(x) \left. \ba
\\
\ea\!\!\!\!\!\!\! \right|_{x=L_{j+1}} = 0, \ \ \ \ \ \ \ \p_x
{\cal S}^{(-)}_{{j+1}}(x) \left. \ba
\\
\ea\!\!\!\!\!\!\! \right|_{x=L_{j+1}} = 1 .
 \ea \label{inibuben}
  \ee
We may immediately make the following two conclusions.

\begin{itemize}

\item
The standard matching of the logarithmic derivatives at every
discontinuity $x = L_j$ finds an elementary though rigorous
guarantee in the obvious rule
 \ben
 c^{(+)}(j)=
 c^{(-)}(j)=c(j),
 \ \ \ \ \ \ \ \ \ \
 d^{(+)}(j)=
 d^{(-)}(j)=d(j),
 \ \ \ \ \ \ \ \ \ \
 j = 0, 1, \ldots, N+1.
 \een
This is the first simplification of our two alternative formulae
(\ref{generalula}) and (\ref{generaluben}) for wave functions.

\item
In each interval $(L_j,L_{j+1})$, {\em any} two-point boundary
conditions $\psi^{(0)}(L_j)=c(j)$, $\psi^{(0)}(L_{j+1})=c(j+1)$
make the solution unique. This statement is equivalent to the two
simple rules
 \be
\ba
 c(j) = c({j+1})\,{\cal C}^{(-)}_{j+1}(L_{j})
  + d({j+1})\, {\cal S}^{(-)}_{j+1}(L_{j})\ ,
\\
c(j+1)= c(j)\,{\cal C}^{(+)}_j(L_{j+1})
  + d(j)\, {\cal S}^{(+)}_j(L_{j+1})
 \ea \label{bcomplul}
 \ee
where, of course, $c(0)=0$ and $c({N+1})=0$ and $j = 0,1, \ldots,
N$. We may summarize that as a net result of our construction we
have to solve just the $2N+2$ homogeneous and linear algebraic
equations for the $2N+2$ unknown parameters.

\end{itemize}

 \noindent
In a marginal comment let us also note that our choice of $c(0)=0$
and $c({N+1})=0$ is dictated by the underlying physics which
requires the most common Dirichlet form of the outer boundary
conditions. In a more formal approach one could also keep in mind
the possible alternative choices of the Neumann conditions
$d(0)=0$ and $d({N+1})=0$ or of the important periodic option for
$c(0)=c({N+1})$ and $d(0)=d({N+1})$ etc. Unfortunately, their
detailed study would already lead us far beyond our present task.

\subsection{Piece-wise polynomial unperturbed potentials
\label{2.4}}

We have seen that the practical use of our nonstandard matching
recipe (\ref{bcomplul}) is a well defined numerical problem. Its
solution requires just the knowledge of the independent sine and
cosine solutions and an evaluation of their values at all the
points of the discontinuities $x=L_j$. It is important to note
that our rigorous matching recipe works with the two independent
bases. This enables us to avoid the more usual but, sometimes,
less comfortable construction of the derivatives of the basis. In
this sense, a slight generalization of this recipe may be easily
applied to all the zero-order problems
  \be
 - {d^2 \over d x^2}\, \psi^{(0)}(x) + V^{(0)}(x)\,
\psi^{(0)}(x) =E^{(0)}\, \psi^{(0)}(x)\  \label{SEzero}
 \ee
with any piece-wise polynomial discontinuous potential
 \be
V^{(0)}(x^\pm) \sim \sum_{l=0}^{p(\pm j)}\, w_l^{\pm}(j)\, (x^\pm
-L_j)^l,\ \ \ \ \ \ \ \ \ x^+ \geq L_j ,\ \ \ \ \ \ \ \ \ x^- \leq
L_j .
 \label{generpo}
 \ee
Locally (i.e., within the ``double" intervals ${\cal J}_j =
(L_{j-1}, L_{j+1})$ such that $ L_{-1}\equiv L_0$ and $L_{N+2}
\equiv L_{N+1}$) we may drop the redundant argument $j$ and
superscripts $^{(0)}$ and search for the exact wave functions in
their respective left and right Taylor series form
 \ben
\psi^{}(x^\pm) \sim \sum_{n=0}^M\ h^{\pm}_n{(j)}(x^\pm -L_{j})^n,
\ \ \ \ \ \ M \to \infty .
 \een
In a purely numerical implementation Hodgson's tests
\cite{Hodgson} confirm the fast convergence of such a ``local"
recipe with $N \to \infty$. In the present perturbation context
the number of discontinuities $x=L_j$ at $j = (0), 1, 2, \ldots,
N, (N+1)$ is fixed and, presumably, very small, $N = {\cal O}(1)$.
Still, in a compactified notation which parallels our previous
$p(\pm j) = 0$ construction we can drop the superscripts $^\pm$
and re-write our local Taylor series as superpositions
 \be \psi(x) = c(j)\,{\cal C}_j(x) + d(j)\, {\cal S}_j(x)\ , \ \ \ \ x
\in (L_{j-1}, L_{j+1}), \ \ \ j =0, 1,  \ldots,N+1.
 \label{general}
 \ee
Their two components ${\cal C}_j(x)={\cal C}(x)$ and ${\cal
S}_j(x)={\cal S}(x)$ are independent solutions of our ordinary
differential Schr\"{o}dinger equation (\ref{SEzero}) again. They
are uniquely determined by their respective cosine-like and
sine-like behaviour at $x=L_j$,
 \be
 \ba
{\cal C}(x) \left. \ba
\\
\ea\!\!\!\!\!\!\! \right|_{x=L_j} = 1, \ \ \ \ \ \ \ \p_x {\cal
C}(x) \left. \ba
\\
\ea\!\!\!\!\!\!\! \right|_{x=L_j} = 0,\\
 {\cal S}(x) \left. \ba
\\
\ea\!\!\!\!\!\!\! \right|_{x=L_j} = 0, \ \ \ \ \ \ \ \p_x {\cal
S}(x) \left. \ba
\\
\ea\!\!\!\!\!\!\! \right|_{x=L_j} = 1 .
 \ea \label{inib} \ee
At the two outer boundaries $L=L_0$ and $R=L_{N+1}$ the current
physical conditions $ \psi(L) =\psi(R) =0$ acquire the most
elementary form $c(0)=0$ and $c({N+1})=0$ or, equivalently,
$c(1)\,{\cal C}_1(L) + d(1)\, {\cal S}_1(L)= 0 $ and $
c({N})\,{\cal C}_N(R) + d({N})\, {\cal S}_N(R)= 0 $. This
parallels again the above $p(\pm j)=0$ special case. The mutual
matchings of the neighboring wave functions are all similar and we
have the final compact set of the physical requirements
\be
\ba c(j)\,{\cal C}_j(L_{j-1}) + d(j)\, {\cal S}_j(L_{j-1})=
c({j-1}),
\\
c(j)\,{\cal C}_j(L_{j+1}) + d(j)\, {\cal S}_j(L_{j+1})= c({j+1}),
\\ \ \ \ \ \ \ \ \ \ \ \ \ \ \
 \ \ \ \ \ \ \ \ j = 1, 2,\ldots,N.
 \ea \label{bcompl}
 \ee
We have to evaluate again the $4N$ input quantities ${\cal
S}_j(L_{j\pm 1})$ and ${\cal C}_j(L_{j\pm 1})$ and solve the
$2N-$dimensional``secular" equation for the arbitrarily normalized
coefficients in the local wave functions (\ref{general}) {\em and}
for the global binding energy. All our piece-wise constant
illustrative examples of subsections \ref{2.1} and \ref{2.2}
re-emerge after the choice of $p(\pm j) = 0$ in potential
(\ref{generpo}) of course.

\section{Matching method for perturbations}

The separate ${\cal O}(\lambda^k)$ components of the perturbed
Schr\"{o}dinger equation have the well known non-homogeneous form
 \be
 - {d^2 \over d x^2}\, \psi^{(k)}(x) + V^{(0)}(x)\,
\psi^{(k)}(x) -E^{(0)}\, \psi^{(k)}(x)
=
\tau^{(k-1)}(x) +E^{(k)}\, \psi^{(0)}(x) \label{SEpert}
 \ee
with
 \ben \tau^{(k-1)}(x) = -V^{(1)}(x)\, \psi^{(k-1)}(x)
+\sum_{j=1}^{k-1} E^{(j)}\, \psi^{(k-1)}(x), \ \ \ \ \ \ \ \  k =
1, 2, \ldots\ . \label{fcert}
 \een
In principle, it defines the $k-$th corrections in terms of their
predecessors $\psi^{(k-1)}(x)$, $\psi^{(k-2)} (x)$, \ldots and
$E^{(k-1)}$, $E^{(k-2)}$, \ldots,  ``compressed" in the
order-dependent right-hand side functions.

\subsection{Local solutions}

The matching method of subsection \ref{2.3} does not use the
(logarithmic) derivatives. This is rendered possible by a certain
redundancy of our construction since domains ${\cal J}_j$ overlap.
We shall now apply the same strategy to the implicit definition
(\ref{SEpert}) of corrections $E^{(k)}\equiv \varepsilon$ and
$\psi^{(k)}(x) = \varphi(\varepsilon,\xi,x)$ split in four terms
{\em locally},
 \be
\psi^{(k)}_j(x)= c^{(k)}(j)\,{\cal C}^{(k)}_j(x) + d^{(k)}(j)\,
{\cal S}^{(k)}_j(x) + \varepsilon\,\omega(x) + \xi^{(k)}_j
\,\psi^{(0)}(x),\ \ \ \ \ j = 1, 2,\ldots,N. \label{eral}
 \ee
This is our key ansatz. Its two free parameters $\varepsilon$ and
$\xi$ should facilitate the matching at the boundaries of ${\cal
J}_j $. Within each of these intervals and at $\varepsilon=0$ and
$\xi=0$ the simplified order-dependent part of our non-homogeneous
differential eq. (\ref{SEpert})
\be
\left[ - {d^2 \over d x^2} + V^{(0)}(x)-E^{(0)} \right]
\,\varphi(0,0,x) = \tau^{(k-1)}\,(x)\  \label{equata} \label{atad}
 \ee
will define all the superpositions of the energy-independent
functions ${\cal C}(x)$ and ${\cal S}(x)$ distinguished (and made
unique) by the respective cosine- and sine-like initial conditions
(\ref{inib}). The third auxiliary function $\omega(x) =
\omega_j(x)$ will be specified as a solution of a simpler,
order-independent equation
\be
\left[ - {d^2 \over d x^2} + V^{(0)}(x)-E^{(0)} \right]
\,\omega(x) = \psi^{(0)}\,(x)\ \label{equabe} \label{atadd}
 \ee
with the different initial conditions
 \ben \omega_j(x) \left. \ba
\\
\ea\!\!\!\!\!\!\! \right|_{x=L_j} = 0, \ \ \ \ \ \ \ \p_x
\omega_j(x) \left. \ba
\\
\ea\!\!\!\!\!\!\! \right|_{x=L_j} = 0\ . \label{qatad}
 \een
The fourth component $\psi^{(0)}(x)$ is known. Its contribution is
weighted by the last parameter $\xi$ chosen to shift the sum
$c^{(k)}(j)+d^{(k)}(j)$ by $c^{(0)}(j)+d^{(0)}(j)$ at each $j$.
Assuming that $c^{(0)}(j)+d^{(0)}(j)\neq 0$ we may re-scale
$c^{(k)}(j)+d^{(k)}(j)=1$ at all $k>0$.

The $k-$ and $j-$dependent variability of $\xi=\xi^{(k)}_j$ does
not violate the validity of our differential Schr\"{o}dinger eq.
(\ref{SEpert}) {\em locally}. The {\em global} matching of its
solutions (\ref{eral}) forms the last step towards an innovated
perturbation method.

\subsection{Global solution in the $k-$th order}

Under the first nontrivial choice of $N=1$ the matching of
perturbation corrections degenerates to the left asymptotic-like
boundary condition at $x=L$ and its right counterpart at $x=R$.
This imposes the two linear algebraic constraints
 \ben \ba
c^{(k)}\,{\cal C}(L) + (1-c^{(k)})\,{\cal S}(L) + \varepsilon\,
\omega(L)=0,\\ c^{(k)}\,{\cal C}(R) + (1-c^{(k)})\,{\cal S}(R)
+\varepsilon\, \omega(R)=0 \ea
 \een
upon the energy $\varepsilon = E^{(k)}$ and coefficient
$c^{(k)}(1)=1-d^{(k)}(1)\equiv X$,
\be
\left[
\begin{array}{cc}
{\cal S}(L) -{\cal C}(L), & -\omega(L)\\ {\cal S}(R) -{\cal C}(R)
,& -\omega(R) \ea \right] \, \left( \ba X\\ \varepsilon \ea
\right)= \left[ \ba {\cal S}(L) \\ {\cal S}(R) \ea \right].
\label{ohon}
 \ee
One has to notice the possible absence of solutions of this system
whenever its determinant vanishes.  Such an apparent paradox just
reflects an {\it a priori} open possiblity of degeneracy of the
unperturbed spectrum.  We only know {\it a posteriori} that the
spectrum of the one-dimensional Sturm-Liouville problem cannot
degenerate at all \cite{Ince}.

Let us abbreviate $ c^{(k)}(j)\,{\cal C}^{(k)}_j(L_i) +
d^{(k)}(j)\, {\cal S}^{(k)}_j(L_i) +\varepsilon\,\omega(L_i)
+\xi_j \,c^{(0)}(i)=\varphi_j(L_i)$ for $N \geq 2$ and, after the
next choice of $N=2$, contemplate the four independent
boundary-and-matching conditions
 \ben
  \ba
   \varphi_1(L_0)=0,\ \ \ \
\ \ \varphi_1(L_2)=\varphi_2(L_2)\ [\
=c^{(k)}(2)+\xi_2\,c^{(0)}(2)],\\ \varphi_2(L_1)=\varphi_1(L_1)\
[\ = c^{(k)}(1)+\xi_1\,c^{(0)}(1)],\
 \ \ \ \ \
\varphi_2(L_3)=0. \ea
 \een
Both the re-normalization parameters enter these equations only in
the form of their difference $Z=\xi_1-\xi_2$. Denoting
$c^{(k)}(1)=1-d^{(k)} (1)\equiv X$ and
$c^{(k)}(2)=1-d^{(k)}(2)\equiv Y$ we get the four linear relations
 \ben \ba
X\,{\cal C}^{(k)}_1(L_0) + (1-X)\, {\cal S}^{(k)}_1(L_0) +
\varepsilon\,\omega(L_0)=0,\\ X\,{\cal C}^{(k)}_1(L_2) + (1-X)\,
{\cal S}^{(k)}_1(L_2) + \varepsilon\,\omega(L_2) + Z \,c^{(0)}(2)
=Y,\\ Y\,{\cal C}^{(k)}_2(L_1) + (1-Y)\, {\cal S}^{(k)}_2(L_1) +
\varepsilon\,\omega(L_1) -Z \,c^{(0)}(1) =X,\\ Y\,{\cal
C}^{(k)}_2(L_3) + (1-Y)\, {\cal S}^{(k)}_2(L_3) +
\varepsilon\,\omega(L_3)=0 \ea
 \een
among the four unknowns $\varepsilon$, $X,\ Y$ and $Z$. This
equation is easily solved by the four-by-four matrix inversion.

At an arbitrary $N$ the general matching plus boundary formula
comprises the $2N$ equations
 \be
\ba X_j\,{\cal C}(L_{j-1}) + (1-X_j)\,{\cal S}(L_{j-1}) +
\varepsilon \omega_j(L_{j-1}) =X_{j-1}+Z_j\,{c}^{(0)}({j-1}),
\\
X_j\,{\cal C}(L_{j+1}) + (1-X_j)\,{\cal S}(L_{j+1}) + \varepsilon
\omega_j(L_{j+1}) =X_{j+1}-Z_{j+1}\, {c}^{(0)}({j+1}),
\\
\ \ \ \ \ \ \ \ \
 \ \ \ \ j = 1,\,2,\, \ldots, N
\ea \label{tilcom}
 \ee
for $2N$ unknowns $\varepsilon$, $X_j=c^{(k)}(j)$ and
$Z_{j+1}=\xi_{j+1}-\xi_{j}$. Our new perturbation prescription is
complete.

\subsection{Illustration}

The detailed implementation of our matching formulae is
straightforward. Its best illustration is provided by the solvable
square well $V^{(0)}(x) = V_{(SW)}(x)$ with the solvable constant
perturbation $V^{(1)}(x) = \Omega$. In this extreme example the
``survival of solvability" facilitates the understanding of
formulae as well as a verification of their quantitative
predictions without any use of a complicated algebra. For the sake
of brevity we shall also pay attention to the $N=1$ recipe in the
first perturbation order only.

\subsubsection{Local wave functions}

In the first step it is easy to extract the particular solution
$\omega^{(part)}(x)= p(x) = ({1/ 2})\, x\,\cos x $ from the
non-homogeneous differential eq. (\ref{atadd}). Its
order-dependent partner eq. (\ref{atad}) looks similar,
 \ben \left[ - {d^2 \over d x^2} + V^{(0)}(x)-E^{(0)}
\right] \,{\cal C}(x) = \tau(x), \ \ \ \ \ \ \ \een $$ \ \ \ \ \ \
\ {\cal C}(x) \left. \ba
\\
\ea\!\!\!\!\!\!\! \right|_{x=X} = 1, \ \ \ \ \ \ \ \ \ \p_x {\cal
C}(x) \left. \ba
\\
\ea\!\!\!\!\!\!\! \right|_{x=X} = 0 $$ \ben \left[ - {d^2 \over d
x^2} + V^{(0)}(x)-E^{(0)} \right] \,{\cal S}(x) = \tau(x), \ \ \ \
\ \ \ \ \ \ \ \ \ \ \ \ \een $$ \ \ \ \ \ \ \ {\cal S}(x) \left.
\ba
\\
\ea\!\!\!\!\!\!\! \right|_{x=X} = 0, \ \ \ \ \ \ \ \p_x {\cal
S}(x) \left. \ba
\\
\ea\!\!\!\!\!\!\! \right|_{x=X} = 1\ ,  $$ and possesses the
similar particular solution $ -\Omega p(x)$. By means of the
Ans{a}tz
 \ben \ba \omega(x)= P\,\sin x + Q\,\cos x + p(x),\\ {\cal
C}(x)= U\,\sin x + A\,\cos x -\Omega\,p(x),\\ {\cal S}(x)= W\,\sin
x + B\,\cos x -\Omega\,p(x)\ \ea
 \een
the initial conditions are easily satisfied by a suitable choice
of the six optional constants $P, Q, U, A, W$ and $B$. After an
elementary trigonometry using the function $q(x) = \p_x p(x) =
(\cos x - x\,\sin x)/2$ and an elementary orthogonal matrix \ben
R(x) = \left(
\begin{array}{cc}
\cos x&\sin x\\-\sin x&\cos x \ea \right)\ \equiv \ [R(-x)]^{-1}
\een we get the result
 \ben \left(
\begin{array}{c}
P\\Q \ea \right)=R(-X)\, \left(
\begin{array}{c}
-p(X)\\-q(X) \ea \right), \een \ben \left(
\begin{array}{c}
A\\U \ea \right)=R(-X)\, \left(
\begin{array}{c}
c+\Omega\,p(X)\\ \Omega\,q(X) \ea \right), \ \ \ \ \ \ \ \left(
\begin{array}{c}
B\\W \ea \right)=R(-X)\, \left(
\begin{array}{c}
\Omega\,p(X)\\d +\Omega\,q(X) \ea \right)\ .
 \een
Our local first-order solution is obtained by fully non-numerical
means.

\subsubsection{Global matching and the energy}

Two-dimensional eq. (\ref{ohon}) represents the physical boundary
conditions at both ends of our interval of coordinates $(L,R)$.
From its matrix elements
 \ben \ba \omega( L)= P,\ \ \ \ \ \ \
\omega( R)= -(P +\pi/2),\ \ \ \ \ \ \ {\cal C}( L)= A,\\ {\cal C}(
R)= \Omega\,\pi/2-A, \ \ \ \ \ \ \  \ \ {\cal S}( L)= B, \ \ \ \ \
\ \ \ \ \ {\cal S}( R)= \Omega\,\pi/2-B\ \ea
 \een it is easy to
deduce the answer $E^{(1)}=\Omega $. This verifies the recipe and
reproduces, incidentally, the exact result.

It is instructive to notice that in the traditional
Rayleigh-Schr\"{o}dinger approach where the value of $\varepsilon$
is evaluated in advance our boundary conditions would be satisfied
automatically. In the present approach the variability of
$\varepsilon$ is admitted breaking, in general, the boundary
conditions
 \ben \left[
\begin{array}{cc}
{\cal S}(L) + \varepsilon\,\omega(L) & {\cal C}(L)+
\varepsilon\,\omega(L)
\\
{\cal S}(R) + \varepsilon\,\omega(R) & {\cal C}(R)+
\varepsilon\,\omega(R) \ea \right] \left( \ba c^{(1)}\\d^{(1)} \ea
\right ) = 0\ .
%\label{fcindi}
 \een
In an apparent paradox the latter equation seems quadratic (but is
linear) in $\varepsilon$. Immediate calculation reveals the
above-mentioned energy correction uniquely,
 \ben
E^{(1)}= \ba \det \left[
\begin{array}{cc}
{\cal S}(L) & {\cal C}(L)
\\
{\cal S}(R) & {\cal C}(R) \ea \right]\\ \hline \det \left[
\begin{array}{cc}
\omega(L) & {\cal S}(L)
\\
\omega(R) & {\cal S}(R) \ea \right] +\det \left[
\begin{array}{cc}
{\cal C}(L) &\omega(L)
\\
{\cal C}(R) &\omega(R) \ea \right] \ea \ = { \Omega \pi \,(B-A)/2
\over \pi\,(B-A)/2}.
 \een
Wave functions may be reproduced in the similar manner.

\section{Summary}

We have seen in Section \ref{2} that for many rectangular
potentials $ V^{(0)}(x)$ the zero-order local wave functions are
superpositions of elementary trigonometric functions $\sin \beta
x$ and $\cos \beta x$ with a real or purely imaginary argument. In
conclusion we should now add that for an arbitrary piece-wise
polynomial perturbation this reduces the construction of
corrections in perturbation series (\ref{7}) to an easy algebraic
exercise.

\subsection{Closed formulae for polynomial perturbations}

Firstly, let us notice that the inhomogeneous term $\tau(x)$ in
eq. (\ref{SEpert}) coincides with a certain superposition of
products $\langle x | k,1\rangle= x^k\, \cos \beta x$ and $\langle
x | k,2 \rangle= x^k\,\sin \beta x$ for polynomial perturbations.
These functions may be denoted and treated as a partitioned basis
$\{\,|k,j\rangle\,\}_{j=1,2}$ with $k = 0, 1,\ldots$. In this
basis the action of the unperturbed differential operator
 \ben \widehat{\cal H}= \left[ - {d^2 \over d
x^2} + V^{(0)}(x)-E^{(0)} \right]
 \een
has a closed explicit form
 \ben \ba
(2\beta)^{-1}\widehat{\cal H}\  x \,\cos \beta x = \sin \beta x,\\
( 4\beta^2)^{-1} \widehat{\cal H}\  (\beta x^2 \cos \beta x-
x\,\sin \beta x) =x\, \sin \beta x,\\ (12\beta^3)^{-1}
\widehat{\cal H}\ (2\beta^2 x^3 \cos \beta x-3 \beta\, x^2\sin
\beta x - 3x\,\cos \beta x) = x^2\sin \beta x,\\ (8\beta^4)^{-1}
\widehat{\cal H}\ (\beta^3 x^4 \cos \beta x- 2\beta^2 x^3\sin
\beta x - 3\beta\, x^2\cos\beta x+ 3\,x\,\sin\beta x) = x^3\sin
\beta x,\\ \ldots\ \ea
 \een
and, {\it mutatis mutandis}, for cosines. Thus, any practical
computation will immediately generalize our previous trivial
constant-perturbation example.

In the constructive proof of the latter relations we firstly
represent the action of our operator $ \widehat{\cal H}$ on each
element of the basis as a superposition of the other basis states.
It is easily shown by explicit differentiation that the
coefficients of these superpositions form an infinite matrix
${\cal Q}$ with the mere three nonzero diagonals. In the second
step, we introduce a two-by-two partitioning of the matrix ${\cal
Q}$ and denote
 \ben {\cal Q}= \left(
\begin{array}{cccccc}
0&\ldots&&&&\\ b_1&0&\ldots&&&\\ c_2&b_2&0&\ldots&&\\
0&c_3&b_3&0&\ldots&\\ 0&0&c_4&b_4&0&\ldots\\ & &&\ddots&\ddots&
\ea \right) .
 \een
The submatrices $b_k=2k\beta\sigma$ and $c_k=-k(k-1)I$ are
elementary and two-dimensional,
  \ben \sigma\ (=
\sqrt{-I})= \left(
\begin{array}{rr}
0&1\\-1&0 \ea \right), \ \ \ \ \ \ \ \ \ I= \left(
\begin{array}{rr}
1&0\\0&1 \ea \right).
 \een
In the third step we verify that the {\em left} inverse ${\cal
Q}^L$ of our singular matrix still exists and has the elementary
form
 \ben
{\cal Q}^L
=
  \left(
\begin{array}{ccccc}
0&b_1^{-1}&0&0&\ldots\\0&
-b_2^{-1}c_2b_1^{-1}&b_2^{-1}&0&\ldots\\0&
b_3^{-1}c_3b_2^{-1}c_2b_1^{-1}& -b_3^{-1}c_3b_2^{-1}
&b_3^{-1}&\ldots\\\vdots&\vdots &\vdots&\vdots&\ddots \ea \right).
 \een
The explicit form of its two-by-two submatrices remains fairly
compact,
 \ben \ba \left({\cal
Q}^L\right)_{nn+1}=-[2b(n+1)]^{-1}{\sigma}, \ \ \ \ \ \
\left({\cal Q}^L\right)_{n+1n+1}=-(2b)^{-2}{I}\\ \left({\cal
Q}^L\right)_{n+2n+1}={(2b)^{-3}(n+2)\sigma}, \ \ \ \ \ \left({\cal
Q}^L\right)_{n+3n+1}={(2b)^{-4}(n+2)(n+3)I}\\ \left({\cal
Q}^L\right)_{n+4n+1}=-{(2b)^{-5}(n+2)(n+3)(n+4)\sigma}, \ \ \ \
\ldots \ , n=0, 1, \ldots \ . \ea
 \een
This completes the proof. The separate rows of the non-partitioned
matrix ${\cal Q}^L$ determine the {particular} solutions of our
fundamental differential equations (\ref{equata}) and
(\ref{equabe}).

\subsection{Generalizations}

Our last observation was extremely pleasant and encouraging. It
immediately implies that for the piece-wise constant unperturbed
potentials our new perturbation construction remains non-numerical
for each polynomial perturbation. In this sense the usual start
from a harmonic oscillator may find here an unexpectedly feasible
methodical alternative even in analyses of continuous models. We
have seen that up to the discontinuities at the lattice points
$x=L_j$, $j = 1, 2, \ldots, N$ the one-dimensional or $s-$wave
functions remained basically non-numerical.

Technical complications may emerge beyond $s-$waves, for
polynomial $V_j^{(0)}(x)$ and for the {nonpolynomial
perturbations}. All these problems may appear quite naturally in
many applications. In such a case both the unperturbed problem
{\em and} the evaluation of corrections become much more
numerical. Still, an implementation of our perturbation recipe
remains virtually unchanged, consisting of the following six
steps.

\begin{itemize}

\item {\bf S 1.}
We solve the {\em unperturbed} differential Schr\"{o}dinger
equation with the appropriate initial conditions (\ref{inib}) in
all the domains ${\cal J}_j$. Their number $N$ is a fixed and,
presumably, very small integer parameter.

\item {\bf S 2.}
We solve the linear algebraic system of the $2\,N$ {\em
homogeneous} equations (\ref{bcompl}). This determines the
unperturbed energy $E^{(0)}$ as well as the unperturbed norms
$c^{(0)}_j$ and $d^{(0)}_j$ of the matched zero order wave
functions.

\item {\bf S 3.}
We solve all the auxiliary initial-value problems (\ref{atadd})
and generate the $N$ functions $\omega(x)=\omega_j(x)$. In
particular, their values  $\omega_j(L_{j\pm 1})$ have to be
computed at the boundaries of all domains.

\item {\bf S 4.}
In the given order $k = 1,2,\ldots$ and in every domain $ {\cal
J}_j$ we specify the ``input" finite sum $\tau_j^{(k-1)}(x)$ and
solve the doublet of the initial value problems (\ref{atad}). This
determines the functions ${\cal C}^{(k)}_j(x)$ and ${\cal
S}^{(k)}_j(x)$ as well as their special values ${\cal
C}^{(k)}_j(L_{j\pm 1})$, ${\cal S}^{(k)}_j(L_{j\pm 1})$.

\item {\bf S 5.}
We solve, finally, our finite set of the $2\,N$ ``effective" or
``model-space" linear algebraic eqs. (\ref{tilcom}). This defines
the $k-$th energy correction $\varepsilon \equiv E^{(k)}$, the $N$
matched norms $ {c}^{(k)}(j)$ and the $N-1$ local re-normalization
parameters $Z_j=\xi_j^{(k)}-\xi_{j-1}^{(k)}$ in the wave
functions.

\item {\bf S 6.}
If needed, we move to the next order $k$ and return to step {\bf S
4.}

\end{itemize}

 \noindent
We may conclude that in the future applications of our new
perturbation prescription its present algebraic (trigonometric)
exemplification may be complemented by some local versions of the
current semi-analytic Taylor series constructions or by the
various discrete (e.g., Runge Kutta) purely numerical
implementations etc.

\section*{Acknowledgements}

Partially supported by the grant Nr. A 1048004 of the Grant Agency
of the Academy of Sciences of the Czech Republic.

%\newpage

\section*{Figure captions}

\noindent Figure 1. Momentum dependence of secular determinant
(\ref{DWAsec}) for four different barriers $H_1=10,\,15,\,20$ and
$25$.

\end{document}